\documentclass[journal=nalefd,manuscript=letter,layout=traditional,email=true,linenumbers]{achemso}

\usepackage{graphicx}   
\usepackage{amsmath}    
\usepackage{enumitem}   
\usepackage{float}      
\usepackage[mathlines]{lineno} 
\usepackage{booktabs}               
\usepackage[hidelinks]{hyperref}    
\usepackage[normalem]{ulem}
\usepackage{array}
\newcolumntype{C}[1]{>{\centering\arraybackslash}m{#1}}
\usepackage{caption}

\usepackage{xcolor}
\definecolor{customred}{RGB}{192,30,40}
\ifdefined\showredtrue
  \DeclareRobustCommand{\red}[1]{{\color{customred}#1}}
\else
  \DeclareRobustCommand{\red}[1]{#1}
\fi

\newcommand{\UWECE}{Department of Electrical and Computer Engineering, University of Wisconsin-Madison, 1415 Engineering Drive, Madison, WI 53706, USA}
\newcommand{\UWMSE}{Department of Materials Science and Engineering, University of Wisconsin-Madison, 1509 University Avenue, Madison, WI 53706, USA}
\newcommand{\UWPHYS}{Department of Physics, University of Wisconsin-Madison, 1150 University Avenue, Madison, WI 53706, USA}
\newcommand{\UNTPhys}{Department of Physics, University of North Texas, 210 Avenue A, Denton, TX 76201, USA}
\newcommand{\ANLCNM}{Center for Nanoscale Materials, Argonne National Laboratory, Lemont, IL 60439, USA}
\newcommand{\Infleqtion}{Infleqtion, Madison, WI, 53703,  USA}

\title{Silicon-on-sapphire metasurfaces generate arrays of dark and bright traps for neutral atoms}

\author{Chengyu~Fang}
\affiliation{\UWECE}

\author{Minjeong~Kim}
\affiliation{\UWECE}

\author{Hongyan~Mei}
\affiliation{\UWECE}

\author{Xuting~Yang}
\affiliation{\UWMSE}

\author{Zhaoning~Yu}
\affiliation{\UWECE}

\author{Yuzhe~Xiao}
\affiliation{\UWECE}
\alsoaffiliation{\UNTPhys}

\author{Sanket~Deshpande}
\affiliation{\UWECE}

\author{Preston~Huft}
\affiliation{\UWPHYS}
\altaffiliation{Present address: QuEra Computing, Inc., 1380 Soldiers Field Road, Boston, MA 02135, USA}

\author{Alan~M.~Dibos}
\affiliation{\ANLCNM}

\author{David~A.~Czaplewski}
\affiliation{\ANLCNM}

\author{Mark~Saffman}
\affiliation{\UWPHYS}
\alsoaffiliation{\Infleqtion}

\author{Jennifer~T.~Choy}
\affiliation{\UWECE}
\alsoaffiliation{\UWMSE}
\alsoaffiliation{\UWPHYS}

\author{Mikhail~A.~Kats}
\affiliation{\UWECE}
\alsoaffiliation{\UWMSE}
\alsoaffiliation{\UWPHYS}
\email{*mkats@wisc.edu}

\begin{document}

\newpage
\begin{abstract}
We demonstrated crystalline silicon–on–sapphire (c-SOS) metasurfaces that convert a Gaussian beam into arrays of complex optical traps, including arrays of optical bottle beams that trap atoms in dark regions interleaved with bright tweezer arrays. The high refractive index and indirect band gap of crystalline silicon makes it possible to design high-resolution near-infrared ($\lambda>700$ nm) metasurfaces that can be manufactured at scale using CMOS-compatible processes. Compared with active components like spatial light modulators (SLMs) that have become widely used to generate trap arrays, metasurfaces provide an indefinitely scalable number of pixels, enabling large arrays of complex traps in a very small form factor, as well as reduced dynamic noise. To design metasurfaces that can generate three-dimensional bottle beams to serve as dark traps, we modified the Gerchberg-Saxton algorithm to enforce complex-amplitude profiles at the focal plane of the metasurface and to optimize the uniformity of the traps across the array. We fabricated and measured c-SOS metasurfaces that convert a Gaussian laser beam into arrays of bright traps, dark traps, and interleaved bright/dark traps.
\end{abstract}

\newpage
\section{INTRODUCTION}
The trapped-neutral-atom array is an emerging platform for quantum information processing, quantum sensing and quantum communications \cite{Kaufman2021}. Arrays of trapped neutral atoms are typically realized using optical tweezers generated by active optical elements, including spatial light modulators (SLMs) \cite{Nogrette2014, Kim2016, Barredo2016, Manetsch2025}, acousto-optic deflectors (AODs) \cite{Endres2016, Bluvstein2022, Bluvstein2024}, and digital micromirror devices (DMDs) \cite{Stuart2018, Wang2020}. While flexible and powerful, these optical devices face several intrinsic limitations. The limited number of pixels (in SLMs and DMDs) or RF tones (in AODs) constrains the upper limit on the number of optical tweezers that can be generated simultaneously;  for instance, neither an N$\times$N DMD nor an N$\times$N SLM can realize an N$\times$N tweezer array because forming the appropriate field distribution around each trap requires multiple degrees of freedom (i.e., pixels in an SLM or DMD), so the system supports far fewer than $N^2$ traps. Also, the large pixel pitch (typically several microns) relative to the trapping laser wavelength ($\sim700$--$1100$~nm) means that demagnification is required to generate diffraction-limited optical traps, complicating the optical setup. Active components can also introduce dynamic noise \cite{Yang2020}, which compromises the stability of quantum systems \cite{Chew2024}. 

An alternative approach is to use passive optical components, for example amplitude masks with a spatial filter and imaging optics that have been used to produce single- and dual-species arrays \cite{Huft2022, Fang2025}, or optical metasurfaces that offer subwavelength phase control through large arrays of engineered nanostructures\cite{Huang2023,Holman2026, Hsu2022, Huang2024, Wang2025}. In other contexts, metasurfaces comprising more than ten billion elements (i.e., pixels) have been demonstrated \cite{Park2024}, suggesting a route for the generation of very large atomic arrays with traps of arbitrary complexity using this technology. \red{To date, however, metasurface demonstrations for atom trapping have been limited to the generation of optical-tweezer arrays. \cite{Huang2023, Huang2024, Holman2026, Wang2025}}

\red{Here, we demonstrate crystalline silicon-on-sapphire (c-SOS) metasurfaces that convert a Gaussian input beam into arrays of complex optical traps: bottle-beam (dark trap) arrays, tweezer (bright trap) arrays, and interleaved bottle-beam and tweezer arrays. Among these, bottle beams are challenging to generate but offer several advantages for quantum applications over optical tweezers: confining atoms at an intensity null reduces photon scattering and enables longer trapping lifetime \cite{Li2012,Piotrowicz2013}, and the small local intensity at the atoms also makes the traps less sensitive to fluctuations of the trapping-laser power. To generate the sophisticated intensity profiles for bottle beams and optimize the uniformity across the array,} we adapted a modified form of the Gerchberg-Saxton (G-S) algorithm \cite{Tao2015} that enforces both amplitude and phase in the image plane of the metasurface. The resulting phase profile was encoded into a c-SOS metasurface, where silicon's high refractive index enables smaller pixel spacing  \cite{Yang2024} compared to metasurfaces with lower-index materials. The resulting metasurfaces, fabricated with CMOS-compatible processes, were used to generate a 7$\times$7 bottle-beam array, a 21$\times$21 bright array, and an interleaved bright(7$\times$7)/dark(6$\times$6) array.



\section{WORKING PRINCIPLE AND DESIGN}

\begin{figure}[H]
    \centering
    \includegraphics[width=0.7\linewidth]{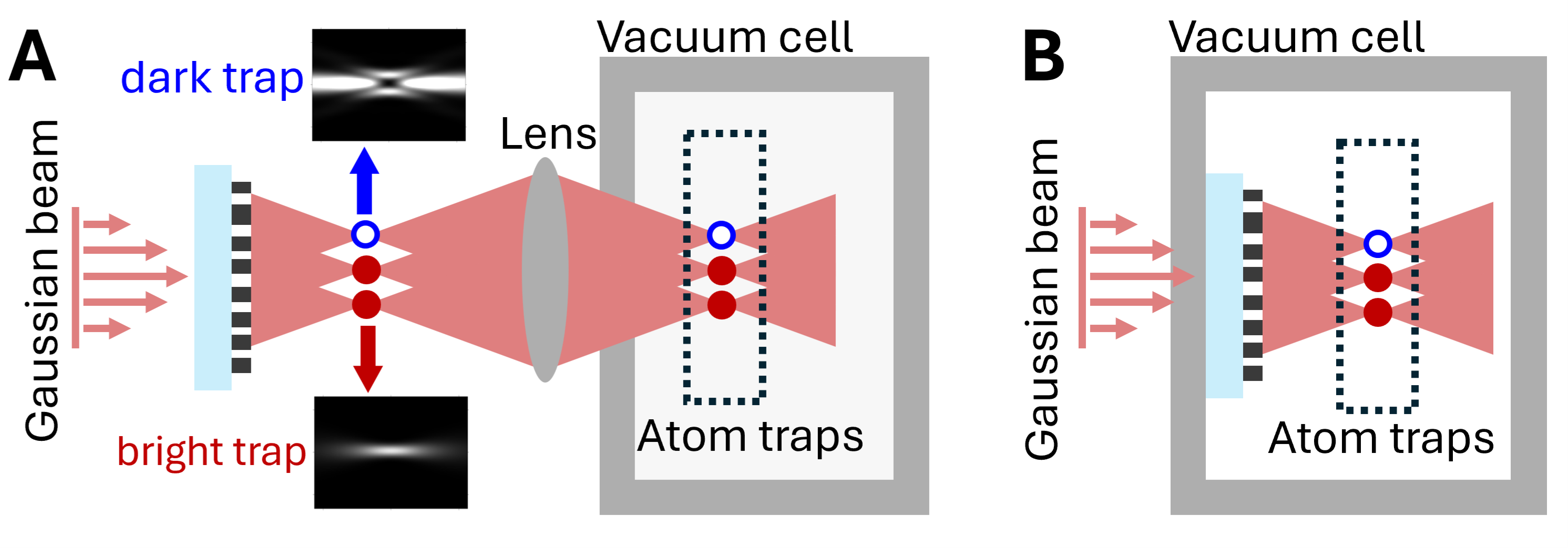}
    \caption{
    Two schemes for positioning a trap-forming metasurface with respect to the vacuum cell.
    \textbf{(A)} Atom-trapping setup with the metasurface placed outside of the vacuum cell. The dark traps (shown in blue) trap the atoms at the intensity minima when the laser frequency is blue-detuned from the atomic resonance, whereas bright traps (shown in red) trap the atoms at the intensity maxima when the laser is red-detuned. Relay optics re-image the intensity profiles into the vacuum cell.
    \textbf{(B)} Schematic of the atom trapping setup with the metasurface integrated inside the vacuum cell. The color convention is the same as (A).}
    \label{fig:figure1}
\end{figure}

Our approach, shown in Fig.~\ref{fig:figure1}, is to send a Gaussian beam through a c-SOS metasurface to form arrays of optical traps at the focal plane. Depending on the target application, the arrays can be configured to be bright tweezer arrays, dark bottle-beam arrays, interleaved bright/dark arrays, or some other desired arrangement. The metasurface can be positioned outside of the vacuum cell  (Fig.~\ref{fig:figure1}(A)) or inside the vacuum cell (Fig.~\ref{fig:figure1}(B)). In the out-of-vacuum configuration, the trapping profile is first generated outside and then re-imaged inside the cell. This approach offers high flexibility, allowing metasurfaces to be swapped without breaking the vacuum and incorporating functions (e.g., beam splitting for fluorescence readout) into the re-imaging optics. Alternatively, in the in-vacuum configuration, the metasurface can be mounted directly inside the vacuum cell, for example as part of the cell wall. This arrangement provides the best integration by eliminating the re-imaging optics, the design of which can be nontrivial due to the need to compensate for the cell wall \cite{Li2020}.

Metasurfaces enable the direct generation of optical traps by applying subwavelength, pixel-by-pixel phase control to the incident Gaussian beam. To guide the design of the phase profiles imparted by the metasurfaces, we need to specify the target optical profiles at the focal plane. A bright trap is typically formed in the focal plane by tightly focusing a Gaussian beam to sub-micron size, whereas a dark trap can be created by overlapping two Gaussian beams with slightly different waists to interfere destructively, generating an intensity minimum at the center \cite{Isenhower2009}.  Arrays of traps are then generated by repeating individual bright or dark traps. For applications in quantum information, the spacing between atoms is typically a few microns, to enable Rydberg interactions \cite{Saffman2010}. 

The metasurface design problem is, then, to take an intensity distribution that forms arrays of bright and/or dark optical traps in the focal plane, and identify the metasurface phase profile that will generate that desired distribution. This looks like a common metasurface design process that is often addressed using one of several versions of the Gerchberg-Saxton (G-S) algorithm; in particular, the G-S algorithm was previously used to design a TiO$_2$ metasurface that generated an array of bright traps \cite{Huang2023, Holman2026}. A conventional G-S algorithm works well for arrays of bright traps because defining the target two-dimensional (2D) amplitude profile at the focal plane is generally sufficient to realize a three-dimensional (3D) tweezer \cite{Huang2023, Holman2026}. 

However, dark traps are more challenging to form, because they require not only intensity minima in the focal plane, but also high-intensity barriers on all sides of the trap in 3D space. Conventional G-S algorithms are not equipped to optimize for such a 3D intensity distribution. Instead of attempting to optimize for a 3D intensity distribution on multiple 2D planes, we instead optimize for the complex field (amplitude and phase) in the focal plane that can define both bright tweezers and bottle beams. Because a bottle beam can be formed by the interference of two Gaussian beams with different waists, the desired amplitude and phase profile in the focal plane can be readily computed. The amplitude profile is the subtraction of two Gaussian profiles with different waists and the phase profile is flat.

\begin{figure}[H]
    \centering
    \includegraphics[width=.7\linewidth]{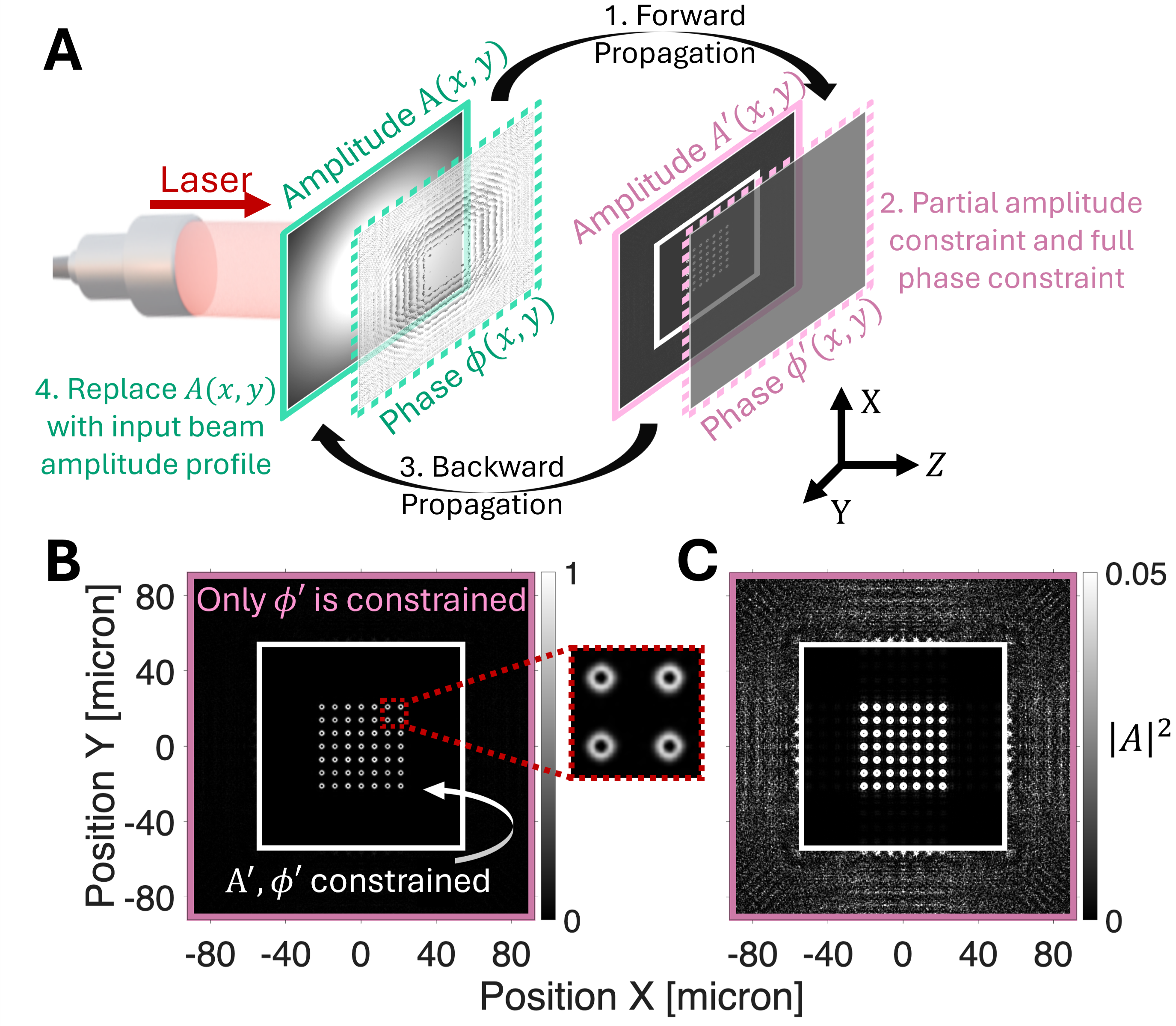}
    \caption{
    Modified Gerchberg-Saxton (G-S) algorithm for generating the bottle-beam array
    \textbf{(A)} Flowchart of the modified G-S algorithm. Each iteration goes through four steps: (1) forward propagation from the metasurface to the image plane; (2) enforcement of both the target amplitude and phase profiles in the fully constrained domain, while only the phase profile is enforced in the partially-constrained domain; (3) backward propagation from the image plane to the metasurface; (4) enforcement of the amplitude profile based on the input (usually Gaussian) beam. When the target amplitude $A'(x,y)$ converges, the resulting metasurface phase $\phi(x,y)$ can be implemented in hardware.
    \textbf{(B)} The simulated intensity profiles in the image plane forming an array of bottle-beam traps, generated using the algorithm in (A). The inset shows a magnified view of several bottle-beam traps. 
    \textbf{(C)} The same profile as (B), but saturated to better show energy leaking into the partially constrained region. }

    \label{fig:figure2}
\end{figure}

We thus use a modified G-S algorithm (illustrated in Fig.~\ref{fig:figure2}(A)) to retrieve the optical phase profile of the metasurface $\phi(x,y)$ that transforms a known input Gaussian beam with an amplitude profile $A(x, y)$, into a target field $A'(x,y)e^{i\phi'(x,y)}$ at the image plane. The algorithm starts with an initial random phase profile of the metasurface with symmetry along $x=0$ and $y=0$ to ensure the resulting optical fields have the same symmetry. The algorithm then proceeds through iterations, each consisting of steps $1-4$:

\begin{enumerate}
    \item \textbf{Forward propagation}: The complex field after passing through the metasurface, $A(x,y)e^{i\phi(x,y)}$ (green in Fig.~\ref{fig:figure2}(A)), is numerically propagated to the image plane via the angular-spectrum method  \cite{Matsushima2009}, resulting in the complex field $A'(x,y)e^{i\phi'(x,y)}$ (magenta in Fig.~\ref{fig:figure2}(A)).

    \item \textbf{Applying image-plane constraints}: In the image plane (magenta in Fig.~\ref{fig:figure2}(A)), we replace the propagated amplitude $A'(x,y)$ in the region demarcated with a white box with a target amplitude. We replace the propagated phase $\phi'(x,y)$ across the entire computational domain. The enforcement of the phase constraint even outside of the region where we have traps ensures that traps near the edge of the array are not distorted. 

    \item \textbf{Backward propagation}: The newly constrained complex field from the image plane is then back-propagated to the metasurface plane via the angular-spectrum method, resulting in a new complex field $A(x,y)e^{i\phi(x,y)}$. The propagation is along $-z$.

    \item \textbf{Applying metasurface-plane constraints}: In the metasurface plane (green in Fig.~\ref{fig:figure2}(A)), the amplitude profile of the newly generated complex-field through backward propagation, $A(x,y)$, is replaced with the input Gaussian beam profile, while the phase $\phi(x,y)$ is retained. This updated complex field then becomes the input for the next iteration.
\end{enumerate}

This loop is repeated until the $A'(x,y)$ converges. The resulting phase profile on the metasurface $\phi(x, y)$ can then be implemented via c-SOS meta-atoms. We demonstrate the results of this optimization process in Fig.~\ref{fig:figure2}(B) , where 7 $\times$ 7 array of bottle-beam traps is formed, surrounded by darkness. To better see the contrast between the fully and partially constrained regions, we saturate the intensity scale in Fig.~\ref{fig:figure2}(C), which shows that some light (unavoidably) leaks into the partially constrained region. 

\section{MATERIALS AND METHODS}

\red{In this work, we implemented our metasurfaces using crystalline silicon-on-sapphire (c-SOS) wafers, which offer several benefits over other materials that have been used for metasurfaces for atom-trap array generation: titanium dioxide (TiO$_2$) \cite{Huang2023, Huang2024} and silicon nitride (SiN$_x$) \cite{Holman2026, Wang2025}. Though by now it is a well-studied material for metasurfaces, TiO$_2$ remains less foundry-compatible because it is challenging to etch \cite{Devlin2016}; fabrication typically relies on conformal atomic layer deposition through a lithographically defined mask followed by planarization \cite{Khorasaninejad2016}. SiN$_x$ is more foundry-compatible, but its lower refractive index is less favorable for meta-atom design and results in the need for higher-aspect-ratio structures \cite{Kamali2018}.

By contrast, crystalline silicon (c-Si) is the most mature material in semiconductor foundries and offers suitable optical properties in the near-infrared range. As shown in Fig.~\ref{fig:si1}, near the D$_2$ resonance wavelengths of $^{87}$Rb (780~nm) and $^{133}$Cs (852~nm), c-Si has a refractive index $\sim$3.6--3.7 and an extinction coefficient less than 0.01. This enables the transmissive metasurface to imprint arbitrary phase profiles on the incident wave without significant optical absorption.
}

\begin{figure}[H] 
    \centering
    \includegraphics[width=1\linewidth]{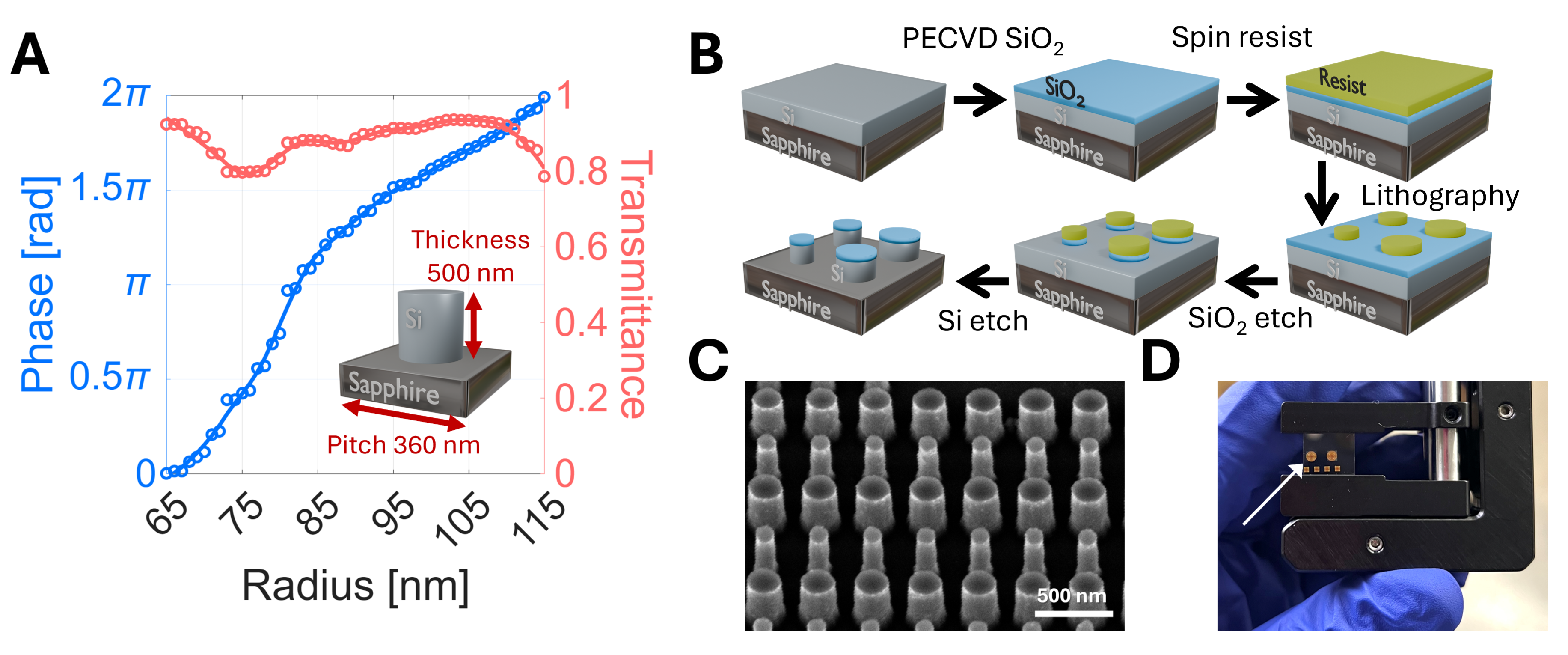}
    \caption{
    Realization of the designed phase profile using silicon-on-sapphire (c-SOS) metasurfaces.
    \textbf{(A)} Simulated transmittance and phase of a single unit cell as a function of the silicon cylinder radius. The height of the cylinder is 500 nm, with a period of 360 nm, on a sapphire substrate. The free-space wavelength of the incident beam in this simulation is $770~\mathrm{nm}$. The transmittance here assumes no reflection at the air-sapphire interface.
    \textbf{(B)} The fabrication processes of the silicon-on-sapphire metasurface. An SiO$_2$ hard mask is first deposited on the silicon-on-sapphire substrate. The patterns are defined by electron-beam lithography and transferred to a hard mask via SiO$_2$ etching. Then silicon etching creates the silicon pillars, which remain covered with the SiO$_2$ mask. Since the refractive index of the amorphous SiO$_2$ is low, the residual SiO$_2$ disk has negligible influence on the performance of the metasurface.
    \textbf{(C)} SEM image of the metasurface on a tilted stage. 
    \textbf{(D)} Photo of the fabricated metasurfaces mounted on a sample holder.}
    
    \label{fig:figure3}
\end{figure}

\red{Our meta-atoms are cylindrical silicon pillars with a pitch of 360~nm, which is less than half the free-space wavelength to suppress higher diffraction orders and avoid spatial aliasing.} The silicon thickness is 500~nm, a choice that balances practical fabrication constraints and the ability to achieve full $0\!\rightarrow\!2\pi$ phase control. With pitch and thickness fixed, we vary the cylinder radius in 1~nm increments to generate a library of complex transmission coefficients, then select a radius range that provides a full $0\!\rightarrow\!2\pi$ phase control while maintaining high transmittance. The corresponding discretized phase and amplitude responses of the meta-atoms are shown in Fig.~\ref{fig:figure3}(A). The metasurface layouts are then generated by mapping the target phase from the modified G-S process (Fig.~\ref{fig:figure2}) at each lattice site to the nearest available phase value, yielding the pillar radius at that site.

The fabrication flow is illustrated in Fig.~\ref{fig:figure3}(B). First, SiO$_2$ is deposited on the c-SOS wafer by plasma-enhanced chemical vapor deposition (PECVD). A positive e-beam resist (ZEP~520A) is spun, soft-baked, and exposed to write the patterns with varying radii that encode the designed phase profile. After developing and short O$_2$ plasma descum, then the pattern is transferred to the SiO$_2$ by reactive ion etching (RIE), to create a hard mask. Next, the c-Si layer is etched by RIE to a target height of 500~nm using HBr/Cl$_2$/O$_2$. The O$_2$ partial pressure is essential to achieve vertical side walls. The SiO$_2$ hard mask is left on the c-Si pillars since it has negligible impact on the optical performance. Scanning electron microscopy (SEM) at an angle of $45^\circ$  was used to verify the fabrication, shown in Fig.~\ref{fig:figure3}(C). A photograph of multiple finished c-SOS metasurfaces on one chip is provided in Fig.~\ref{fig:figure3}(D). We note that the optical function of the metasurfaces is robust to reasonable fabrication errors, such as non-vertical sidewalls resulting from imperfect etching. As an example, we simulated the performance of the dark-trap array metasurfaces using hourglass-shaped pillars, as shown in Fig.~\ref{fig:si5}.

\section{RESULTS}
\subsection{Intensity profile of single dark trap}

\begin{figure}[H] 
    \centering
    \includegraphics[width=.6\linewidth]{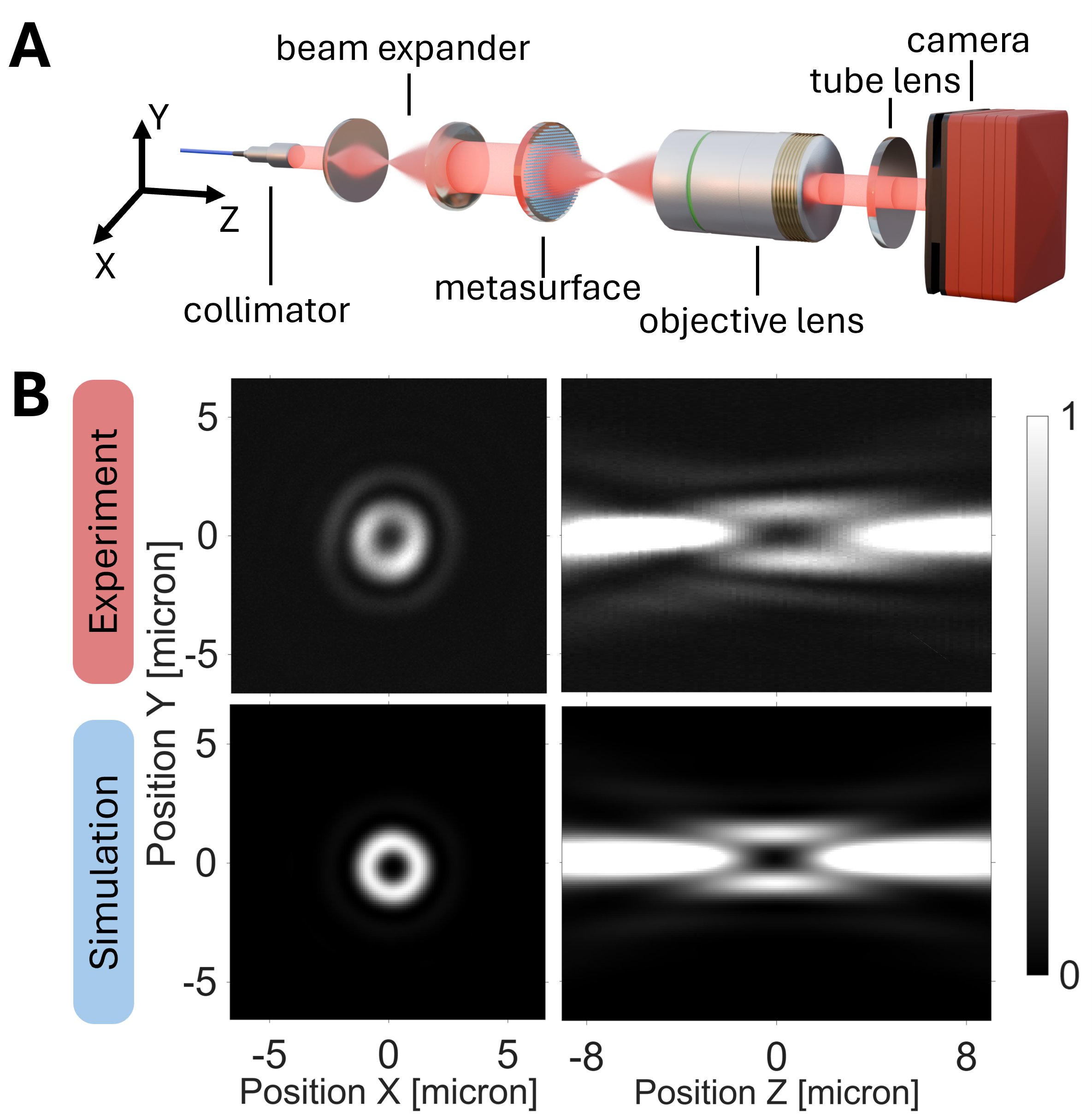}
    \caption{
    Generation and measurement of a single dark trap. 
    \textbf{(A)} Experimental setup for imaging the intensity profile after the metasurface. A Gaussian beam at $780~\mathrm{nm}$ is launched from a single-mode fiber with a collimator and expanded to the desired beam waist. The metasurface is mounted on a motorized XYZ stage combined with a tip-tilt stage. The intensity profiles of the optical traps are acquired with a microscope, while scanning the metasurface along the Z-direction.
    \textbf{(B)} Experimental (top row) and simulated (bottom row) intensity profiles. Left plots show the XY intensity profiles in the focal plane at $Z = 350~\mu\mathrm{m}$, and right plots show the YZ intensity profiles at $X = 0~\mu\mathrm{m}$. All profiles are normalized to the maximum intensity of the corresponding XY intensity profile.}
    \label{fig:figure4}
\end{figure}

We first designed and fabricated a metasurface for generation of a single dark trap (Fig.~\ref{fig:figure4}), similar to one of the designs using accelerating beams \cite{Xiao2021}. The specific target dark trap consists of the interference of two Gaussian beams with waists $w_1=1.33~\mu\mathrm{m}$ and $w_2=0.7~\mu\mathrm{m}$, such that there is total destructive interference at the beam waist. As a result, the target amplitude profile is the subtraction of these two Gaussian profiles at the beam waists and the phase profile is flat.

The optical characterization setup is shown in Fig.~\ref{fig:figure4}(A). A $780~\mathrm{nm}$ Gaussian beam from a single-mode fiber was expanded to the designed input waist $w_0 = 0.13~\mathrm{mm}$ before illuminating the metasurface. The metasurface was placed at the beam waist of the input Gaussian beam and mounted on a motorized XYZ and tip-tilt stages. These stages were carefully adjusted to ensure that the metasurface center was aligned with the Gaussian beam center, and the beam was strictly perpendicular to the metasurface. The dark trap forms at a working distance of $0.35~\mathrm{mm}$. To measure the intensity profile, a microscope is built on the other side of the trap with an objective lens ($f_1 = 4.5~\mathrm{mm}$, NA = 0.65), a tube lens ($f_2 = 50~\mathrm{mm}$), and a monochrome camera (Allied Vision Alvium 1800 U-501m NIR). The metasurface is translated along the optical axis (Z) to acquire a full 3D map of the intensity. Strictly speaking, we should translate the microscope to keep the input beam unchanged on the metasurface; however, because the beam's Rayleigh length is much longer than the scanning range, we instead translate the metasurface, which is simpler than moving the entire microscope.

Fig.~\ref{fig:figure4}(B) compares measured and simulated intensity profiles, which are both normalized to the peak intensity in the focal plane. The measured XY profile shows the expected ``donut'' shape in the focal plane, with a low-intensity center. The measured YZ profile at $X = 0$ shows a bottle-beam geometry. Because of the camera's limited dynamic range, we do not report a reliable trap depth value here. Nonetheless, the close agreement between the experiment and simulation results confirms that the designed and fabricated metasurface generates the dark trap as intended.

\subsection{Intensity profiles for one- and two-species arrays}

\begin{figure}[H]
    \centering
    \includegraphics[width=1\linewidth, trim={15pt 2pt 2pt 15pt}, clip]{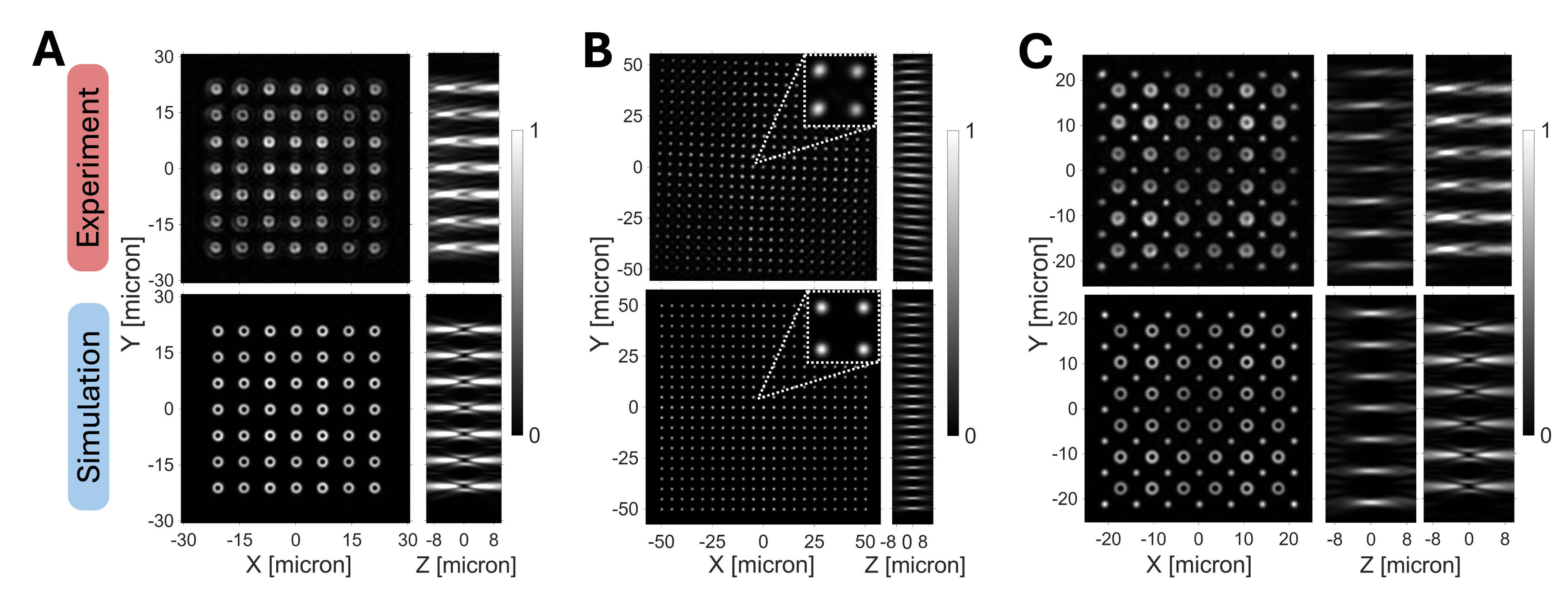}
    \caption{
    Experimental and simulated intensity distributions for dark, bright, and dual-species trap arrays.
    Experiments (top row) and simulations (bottom row). The left plots of each panel show the XY intensity profiles in the focal plane, at $Z = 350~\mu\mathrm{m}$. The right plots of each panel show the YZ intensity profiles across trap sites. In panels (A) and (B), the profiles are along $X = 0~\mu\mathrm{m}$; for panel (C), the profiles are along $X = 0~\mu\mathrm{m}$ (bright traps) and $X = 3.5~\mu\mathrm{m}$ (dark traps). All profiles are normalized to the maximum intensity of the corresponding XY intensity profile, with line profiles shown in Fig.~\ref{fig:si6}.
    \textbf{(A)} 7$\times$7 dark-trap array
    \red{\textbf{(B)} 21$\times$21 bright-trap array. The insets in the XY plots show magnified views of the four traps at the center.}
    \textbf{(C)} 7$\times$7 bright-trap array interleaved with 6$\times$6 dark-trap array.}
    \label{fig:figure5}
\end{figure}

We further demonstrate metasurface-generated optical trap arrays in both single-species and dual-species configurations. The target amplitude profile of an array is generated by placing copies of the single-trap target amplitude profile at the desired lattice sites. We designed arrays for single-species traps (all bright and all dark) or a mixture of bright and dark traps. The target phase profile is set to be flat across the entire trapping plane, to generate the dark traps and improve the uniformity across the array. In Fig.~\ref{fig:figure5}, we compare the experimentally measured and simulated intensity distributions for (A) a dark-trap array, (B) a bright-trap array, and (C) an interleaved dual-species array that combines both types of traps, using the same optical setup in Fig.~\ref{fig:figure4}(A). 

For the single-species arrays shown in Fig.~\ref{fig:figure5}(A, B), the metasurfaces generate dark (blue-detuned) and bright (red-detuned) traps, respectively. We selected modest array sizes and lattice periods to make it easy to image the arrays, but the approach shown here can be used for much larger arrays with spacing down to about 3 $\mu \mathrm{m}$ (or even smaller, with higher NA), though the bright traps can be somewhat closer together than the dark traps. 

The dark-trap array in Fig.~\ref{fig:figure5}(A) consists of a $7\times7$ periodic lattice with a period of $7~\mu$m, designed using the same parameters as the single dark trap in Fig.~\ref{fig:figure4}(B). The bright-trap array in Fig.~\ref{fig:figure5}(B) consists of $21\times21$ traps with a period of $5~\mu\mathrm{m}$; each bright trap has a Gaussian profile with a waist of $0.94~\mu\mathrm{m}$, which matches the trap size of dark traps. The measured XY intensity profiles reveal periodic arrays of donut-shaped dark traps in Fig.~\ref{fig:figure5}(A) and bright focal spots in Fig.~\ref{fig:figure5}(B), while the YZ cross sections at $X=0$ confirm strong axial confinement in both cases. Excellent agreement was obtained between the simulations and the measured intensity profiles.

We also designed a metasurface capable of simultaneously generating interleaved bright and dark traps (Fig.~\ref{fig:figure5}(C)), an arrangement of interest for two-species architectures for quantum error correction \cite{Anand2024, Petrosyan2024}. In this configuration, the trapping laser wavelength would be chosen to be between the resonance wavelengths of the two atomic species, making it red-detuned for one species and blue-detuned for the other. The focal plane of the metasurface in Fig.~\ref{fig:figure5}(C) contains $7\times7$ bright traps interleaved with a $6\times6$ dark-trap array, forming a checkerboard pattern, as shown in the XY intensity profile. The YZ cross section at $X = 0~\mu\mathrm{m}$ displays the bright-trap array, while the cross section at $X=2.5~\mu\mathrm{m}$ corresponds to the dark-trap array. 

\section{DISCUSSION}

The results in Fig.~\ref{fig:figure5} show that the metasurfaces can generate trap-arrays for one or two atomic species, using a single laser and with no additional optical components. To quantitatively evaluate the performance of these metasurface-generated trap arrays, we calculated the trap depth at each individual trapping site and the standard deviation across the entire array for dark- and bright-trap arrays. In this analysis, we assume the input trapping laser is a Gaussian beam with a waist of roughly $130~\mu\mathrm{m}$ at a power of $1~\mathrm{W}$ incident on the metasurface. While the laser wavelength varies depending on the specific application settings, such as the atomic species or the availability of high-power laser, we use $\lambda = 780~\mathrm{nm}$ as an example for the following calculations, since the same metasurface design performs well at nearby wavelengths (see, e.g., Fig.~\ref{fig:si4}).

\begin{figure}[H] 
    \centering
    \includegraphics[width=1\linewidth]{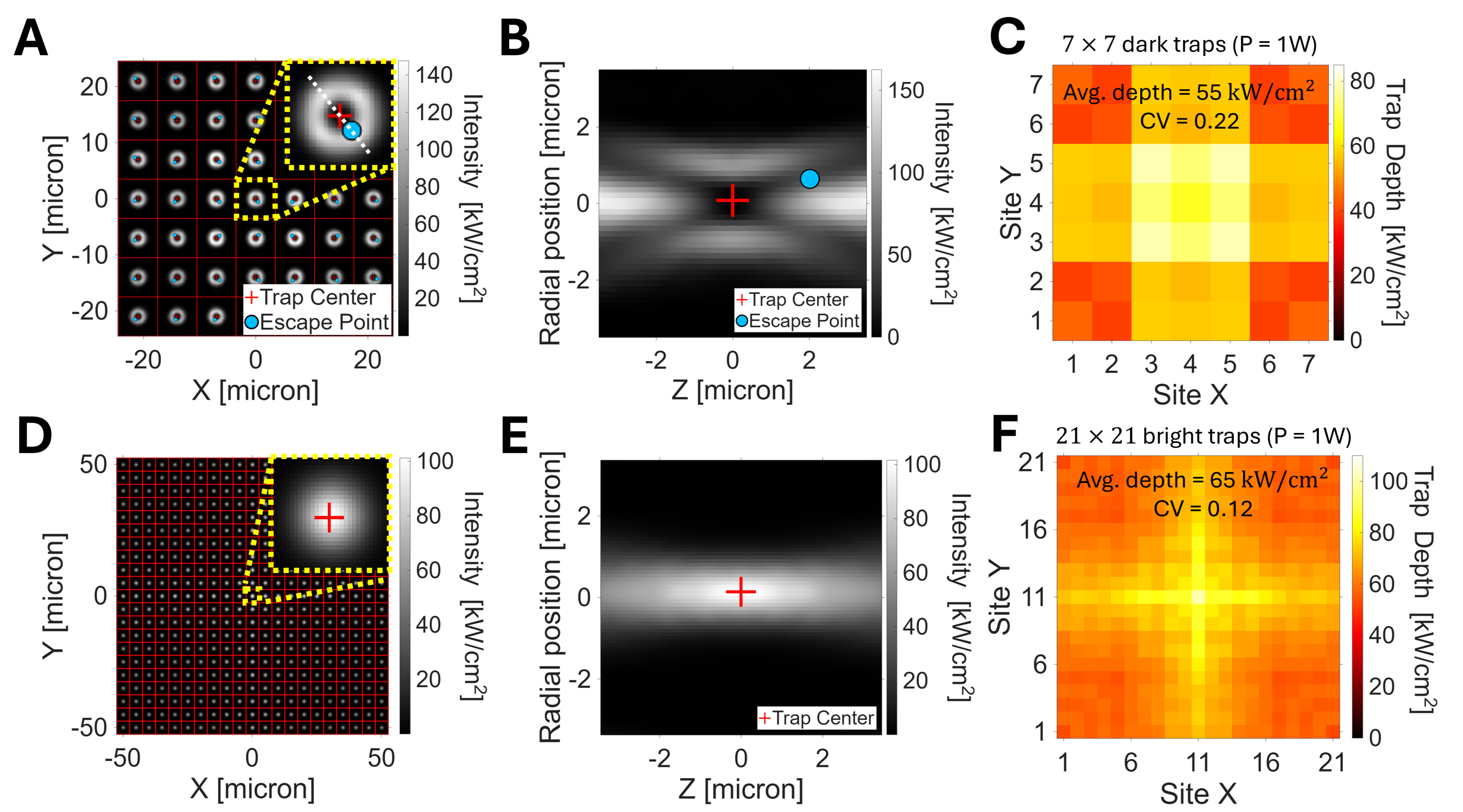}
    \caption{Trap depth analysis of dark (A--C) and bright (D--F) trap arrays. 
    \textbf{(A)} XY intensity profile for a $7 \times 7$ dark-trap array. The inset shows a magnified view of a single dark trap, where the red cross marks the trap center (intensity minimum) and the blue circle indicates the calculated escape point. Note that in the central trap sites, there are multiple escape points due to symmetry, and we plot the one found by our algorithm.
    \textbf{(B)} ``Radial'' cross-section across a single dark trap showing 3D confinement, where the specific cut is along the direction given by the trap center and the projection of the escape point on the X-Y plane, as shown in the inset of (A). 
    \textbf{(C)} Trap-depth distribution heatmap for the $7 \times 7$ dark array at $1~\mathrm{W}$ input. 
    \textbf{(D)} XY intensity profile for a $21 \times 21$ bright-trap array. The inset shows a magnified view of the central dark trap, where the red cross marks the trap center (intensity maximum).
    \textbf{(E)} Radial cross-section of the central bright trap, showing 3D confinement.
    \textbf{(F)} Trap depth distribution heatmap for the $21 \times 21$ bright array at $1~\mathrm{W}$ input.}
    \label{fig:figure6}
\end{figure}

\red{We use the trap depth at a given input power as our figure of merit, rather than the efficiency of light delivery to the region of the traps. A higher efficiency does not necessarily translate to better dark traps since deformation of the dark traps can cause leakage of atoms.} For the dark traps, the trap intensity depth is defined as the intensity difference between the minimum at the trap center and the nearby ``escape point'', as illustrated in Fig.~\ref{fig:figure6}(A, B). The escape point is the location where an atom would be the most likely to exit the trap. To identify this point, we use a modified Dijkstra's algorithm \cite{Pollack1960} on the three-dimensional intensity distribution to search for the escape path that minimizes the maximum potential barrier along the route. The escape point is then defined as the maximum intensity along this optimal path. We note that the escape points in the dark traps are not located in the X-Y plane; to visualize the exact location, in Fig.~\ref{fig:figure6}(B), we plot the cross-section intensity profile across the trap, taking the cut that crosses the center point and the calculated escape point. Note that the central traps have several equivalent escape points due to symmetry, and our algorithm finds one of them. Using this definition, we calculate the trap depths for every site across the $7\times7$ dark-trap array, resulting in the trap-depth distribution shown in Fig.~\ref{fig:figure6}(C). The average trap depth is $55~\mathrm{kW/cm^2}$ with coefficient of variation (CV) of 0.22, where CV is defined as the standard deviation divided by the mean. 

For the bright traps, we define the trap intensity depth as the difference between the peak intensity at the trap center and the surrounding background intensity, as shown in Fig.~\ref{fig:figure6}(D,E). Because the background intensity is negligible compared to the peak intensity, the trap depth can be just treated as the peak intensity. Using this definition, we calculate the trap depths for every site across the $21\times21$ bright-trap array, giving us the trap-depth distribution shown in Fig.~\ref{fig:figure6}(F). The average trap depth is $65~\mathrm{kW/cm^2}$ with CV = 0.12. 

We found that it is significantly more energy-efficient to form bright traps than dark traps. For a given amount of laser power and designing a bright-trap array and a dark-trap array with similar intensity trap depths, the bright-trap array would have approximately nine-times more traps than the dark-trap array. This increased power consumption for dark traps is because the optical field must form a bottle-like distribution that surrounds the trap center, rather than being focused to the trap center as in bright traps. This bottle-like distribution takes a larger volume and thus leads to a lower intensity contrast. While bright traps are more power-efficient, dark traps may still be preferred for applications requiring long coherence time and minimal scattering rate \cite{Saffman2010, Katori2011}. We also investigated the sensitivity of the metasurfaces to input beam misalignment for dark-trap arrays. Our simulations show that the metasurfaces are robust to moderate misalignment in the incident angle (Fig.~\ref{fig:si2}) and the beam center (Fig.~\ref{fig:si3}).

In summary, we experimentally demonstrated a CMOS-compatible silicon-based metasurface platform for atom-trapping applications using near-infrared lasers. We realized metasurfaces that generate arrays of bright traps, dark traps, and interleaved dark and bright arrays, using a single input laser. Compared to approaches based on active beam-shaping devices (e.g., AODs), our metasurfaces are passive, eliminating noise due to active components, and potentially improving trap stability. The metasurfaces and the resulting trap arrays can be scaled to nearly arbitrary sizes (e.g., see Fig.~\ref{fig:si7} and \ref{fig:si8}), limited primarily by fabrication time if using serial techniques such as electron-beam lithography, and the availability of high-power lasers. The metasurfaces are also highly compact, and thus can be integrated inside the vacuum cell, supporting miniaturized quantum setups. This paper positions crystalline-silicon-based metasurfaces as a scalable tool for atom-based quantum technologies. \red{Looking ahead, we envision metasurfaces as an element of a hybrid architecture for neutral-atom platforms. In such an architecture, a passive metasurface provides a background trapping array with complex trapping profiles and/or large number of trapping sites, while active optical components (e.g., AODs) offer dynamic functionalities such as atom rearrangement.}


\section*{Competing Interests}
M.S. is a shareholder in Infleqtion. All other authors declare they have no competing interests.

\section*{Acknowledgments}
The UW-Madison portion of this work is supported by the U.S. Department of Energy Office of Science National Quantum Information Science Research Centers as part of the Q-NEXT center (majority), and by the National Science Foundation under Award 2016136 for the QLCI center Hybrid Quantum Architectures and Networks. Work performed at the Center for Nanoscale Materials, a U.S. Department of Energy Office of Science User Facility, was supported by the U.S. DOE, Office of Basic Energy Sciences, under Contract No. DE-AC02-06CH11357. X.Y. and J.T.C. were supported by the Office of Naval Research under Grant No. N00014-20-1-2598

\section*{Author Contributions}
Conceptualization: C.F., M.K., Z.Y., Y.X., S.D., P.H., M.S., and M.A.K.
Calculations and simulations: C.F., M.K, Y.X..
Experimental realization: C.F., M.K, H.M., X.Y., A.M.D., D.A.C, J.T.C.
Data analysis: C.F, M.K., M.A.K.
Project administration: M.A.K., J.T.C, and M.S. 
Writing original draft: C.F. and M.A.K.
Review and editing: all authors


\bibliography{references}


\setcounter{figure}{0}
\setcounter{section}{0}
\setcounter{table}{0}
\renewcommand{\figurename}{Figure}
\renewcommand{\thesection}{S\arabic{section}}
\renewcommand{\thefigure}{S\arabic{figure}}
\renewcommand{\thetable}{S\arabic{table}}

\newpage
\section*{\Large Supplementary Materials}

\begin{enumerate}
    \item[] \nameref{sec:si1}
    \item[] \nameref{sec:si2}
    \item[] \nameref{sec:si3}
    \item[] \nameref{sec:si4}
    \item[] \nameref{sec:si5}
    \item[] \nameref{sec:si6}
    \item[] \nameref{sec:si7}
    \item[] \nameref{sec:si8}
\end{enumerate}


\newpage
\section{1. Optical properties of crystalline silicon on a sapphire substrate}
\label{sec:si1}

\begin{figure}[H]
    \centering
    \includegraphics[width=0.5\linewidth]{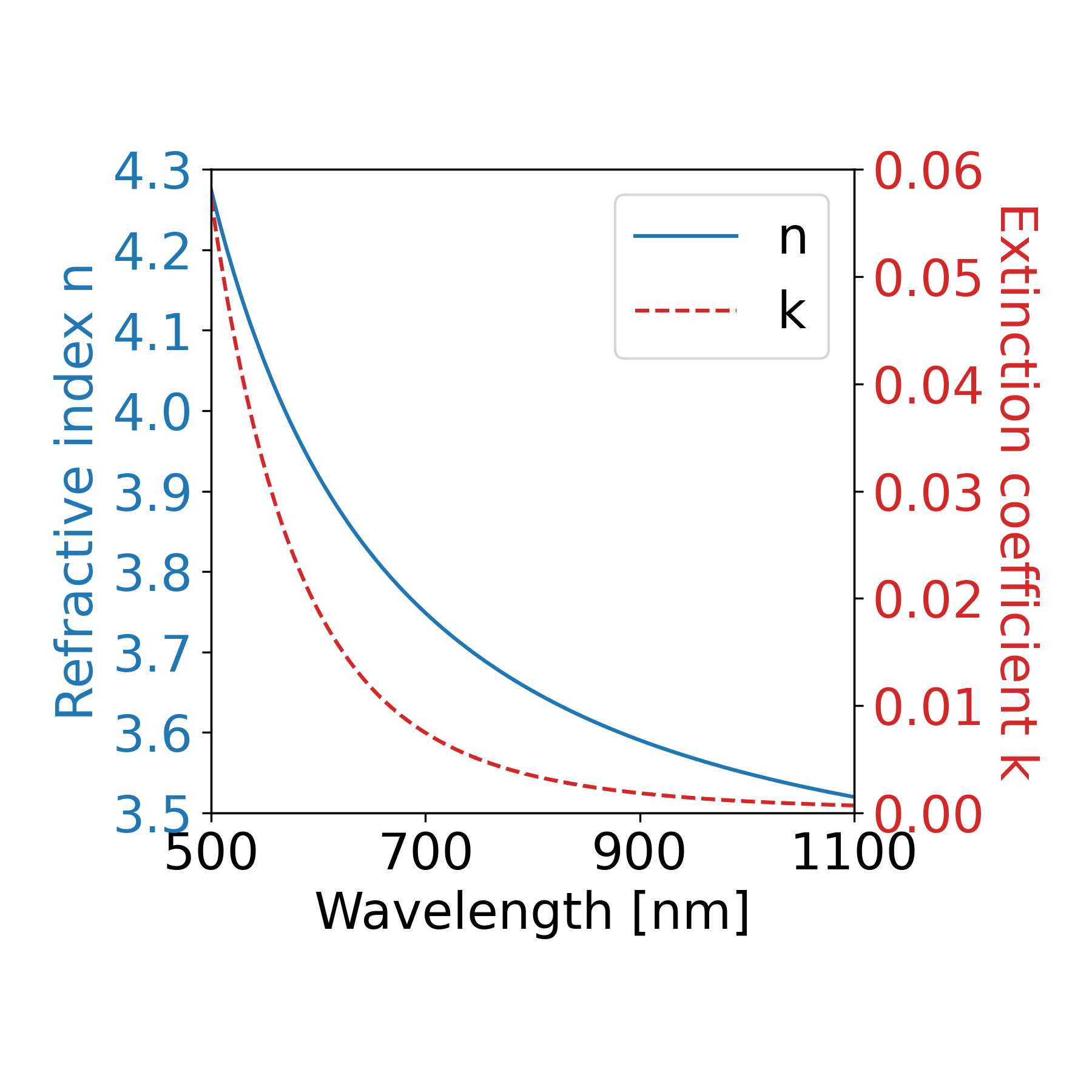} 
    \caption{
    Wavelength-dependent complex refractive index $n + ik$, measured using spectroscopic ellipsometry. The real part $n$ (blue solid, left $y$-axis) is relatively high ($\sim$3.5--4.3) from 500 to 1100~nm, 
        while the extinction coefficient $k$ (red dashed, right $y$-axis) remains low 
        ($<0.01$ for wavelengths longer than $\sim$700~nm). 
    }
    \label{fig:si1}
\end{figure}

\newpage
\section{2. Effects of an off-angle incident Gaussian beam}
\label{sec:si2}
\begin{figure}[H]
    \centering
    \includegraphics[width=.8\linewidth]{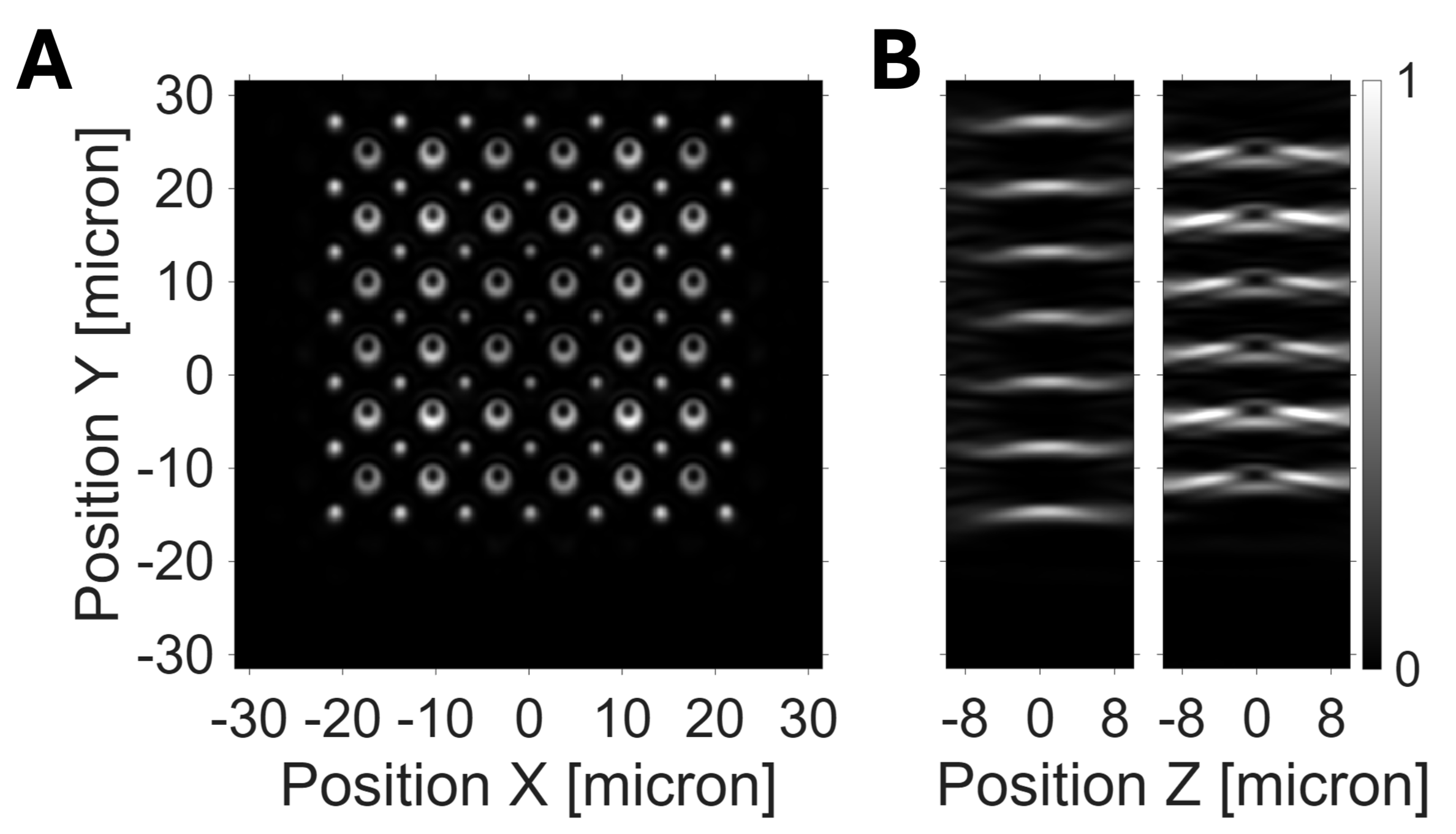} 
    \caption{
    Simulated intensity profile of a two-species trap array generated by a metasurface, assuming the incident beam is $1^\circ$ off the normal angle. The metasurface design is the same as the one in Fig. \ref{fig:figure5}(C).
    \textbf{(A)} XY intensity profile in the focal plane at $Z = 350~\mu\mathrm{m}$.
    \textbf{(B)} YZ intensity profiles along $X = 0~\mu\mathrm{m}$ (bright traps) and $X = 3.5~\mu\mathrm{m}$ (dark traps).
    }
    \label{fig:si2}
\end{figure}

\newpage
\section{3. Effects of an off-center incident Gaussian beam}
\label{sec:si3}
\begin{figure}[H]
    \centering
    \includegraphics[width=.8\linewidth]{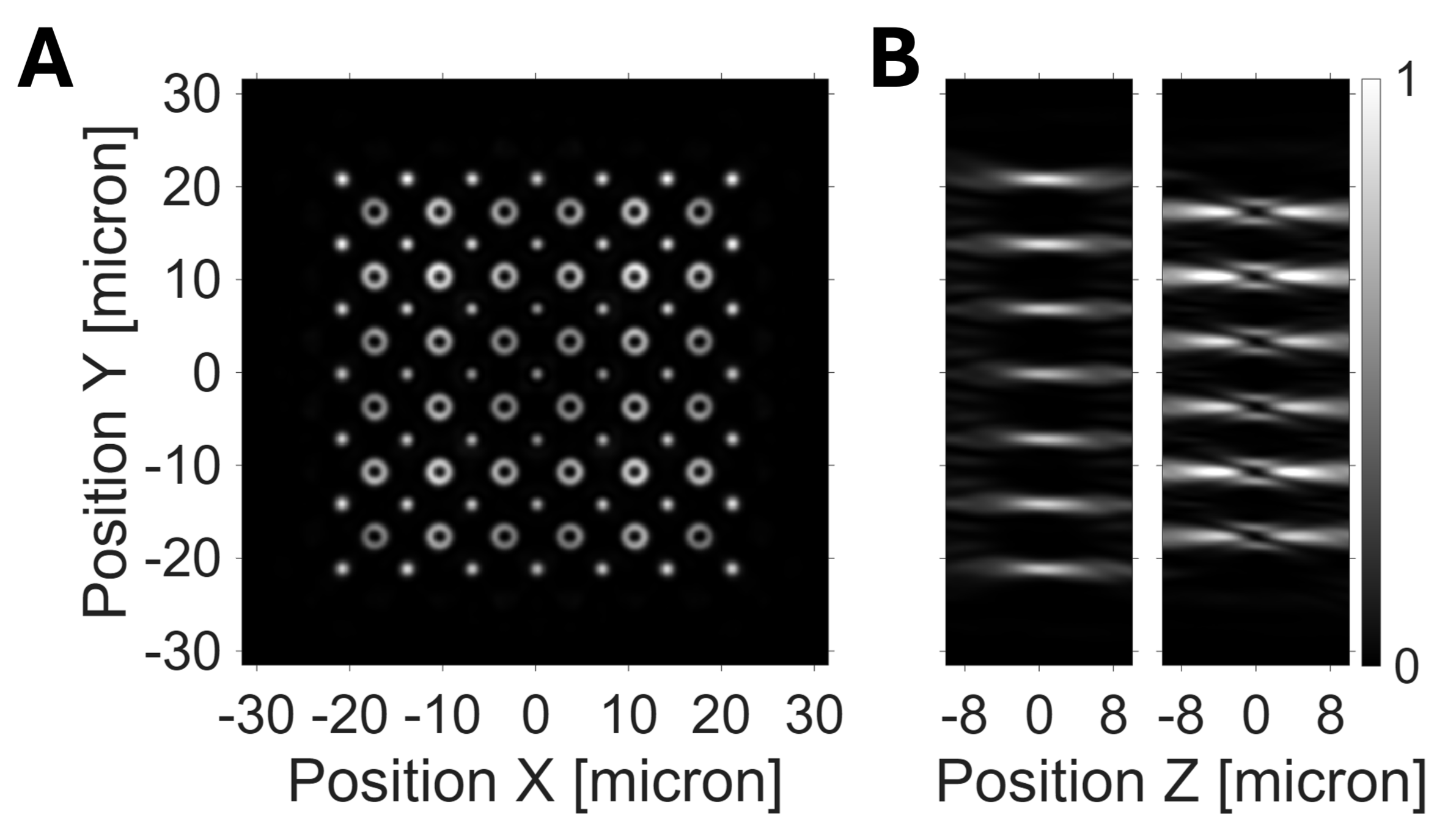} 
    \caption{
    Simulated intensity profile of a two-species trap array generated by a metasurface, assuming the incident beam is $20~\mu\mathrm{m}$ off the center. The metasurface design is the same as the one in Fig. \ref{fig:figure5}(C).
    \textbf{(A)} XY intensity profile in the focal plane at $Z = 350~\mu\mathrm{m}$.
    \textbf{(B)} YZ intensity profiles along $X = 0~\mu\mathrm{m}$ (bright traps) and $X = 3.5~\mu\mathrm{m}$ (dark traps).
    }
    \label{fig:si3}
\end{figure}

\newpage
\section{4. Performance of the dark-trap-array metasurface at a different input wavelength}
\label{sec:si4}
\begin{figure}[H]
    \centering
    \includegraphics[width=1\linewidth]{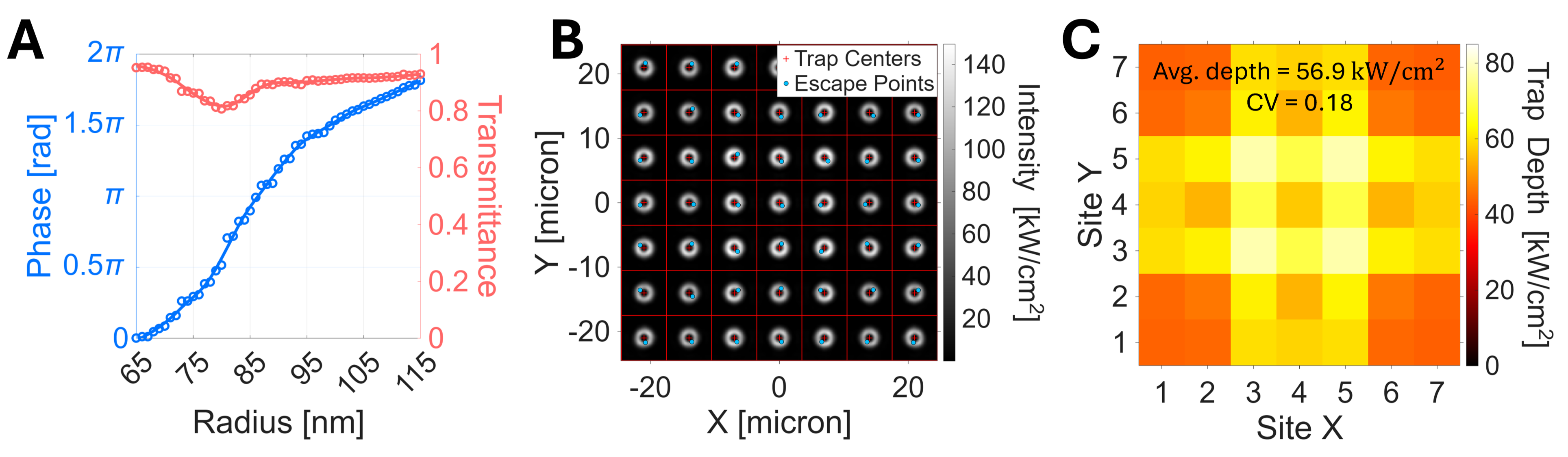} 
    \caption{Performance of the dark-trap metasurface designed for an input wavelength of $\lambda = 770~\mathrm{nm}$ but operating at $\lambda = 800~\mathrm{nm}$. 
    \textbf{(A)} Simulated transmittance and phase of a single unit cell as a function of the silicon cylinder radius. The height of the cylinders is 500 nm, with a period of 360 nm, on a sapphire substrate. The free-space wavelength of the incident beam in this simulation is 800 nm.
    \textbf{(B)} Simulated XY intensity profile of the $7\times7$ dark-trap array in the focal plane. 
    \textbf{(C)} Trap depth distribution heatmap for the array at $1~\mathrm{W}$ input power.}
    \label{fig:si4}
\end{figure}

\newpage
\red{
\section{5. Performance of the dark-trap-array metasurface considering fabrication errors}

In the processes of metasurface fabrication, the shape of the silicon pillar might differ from the ideal cylindrical shape. As an example, due to imperfect etching processes, the sidewall of the pillars may not be vertical. Here, we simulated the performance of the dark-trap array metasurface with hourglass-shaped sidewalls, as shown in Fig.~\ref{fig:si5}. We observe that the metasurface still generates a dark-trap array but with a slightly lower average trapping depth and similar CV.
}

\red{
\label{sec:si5}
\begin{figure}[H]
    \ifdefined\showredtrue
      \captionsetup{font={color=customred}, labelfont={color=customred}}
    \fi
    \centering
    \includegraphics[width=\linewidth]{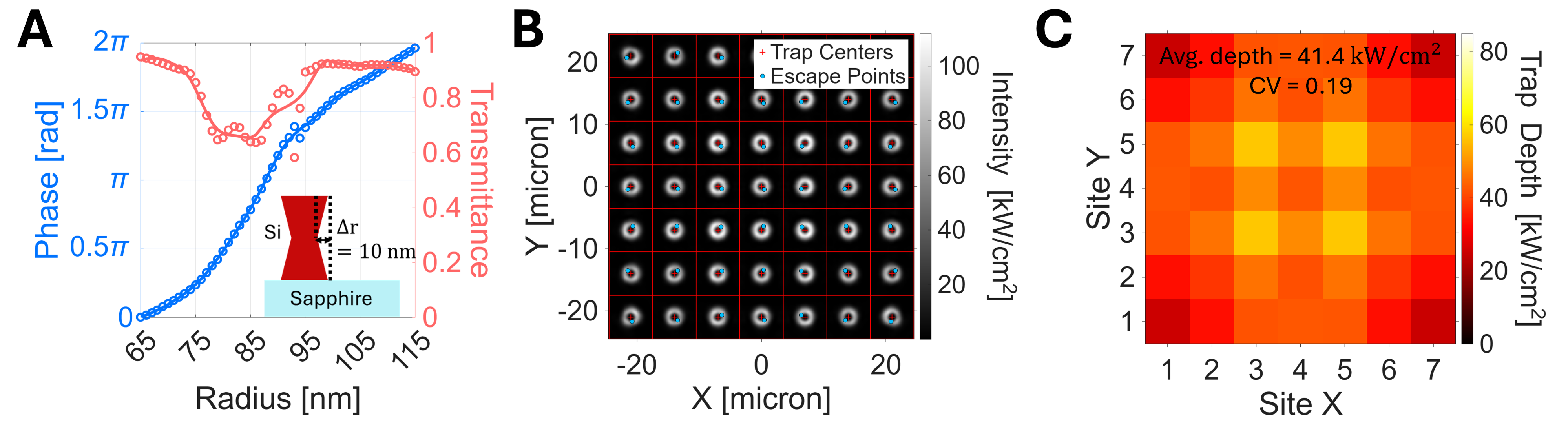}
    \caption{
    \textbf{(A)} Simulated transmittance and phase response of a single unit cell as a function of the silicon pillar radius. The pillars are hourglass-shaped to represent fabrication imperfections, with a radius waist at the middle that is 10 nm smaller than the pillar radius.
    \textbf{(B)} Simulated XY intensity profile of the $7 \times 7$ dark-trap array in the focal plane.
    \textbf{(C)} Trap depth distribution heatmap for the array at $1~\mathrm{W}$ input power.
    }
    \label{fig:si5}
\end{figure}
}

\newpage
\section{6. Intensity line cuts through the experimental intensity profiles}
\label{sec:si6}
\begin{figure}[H]
    \centering
    \includegraphics[width=\linewidth]{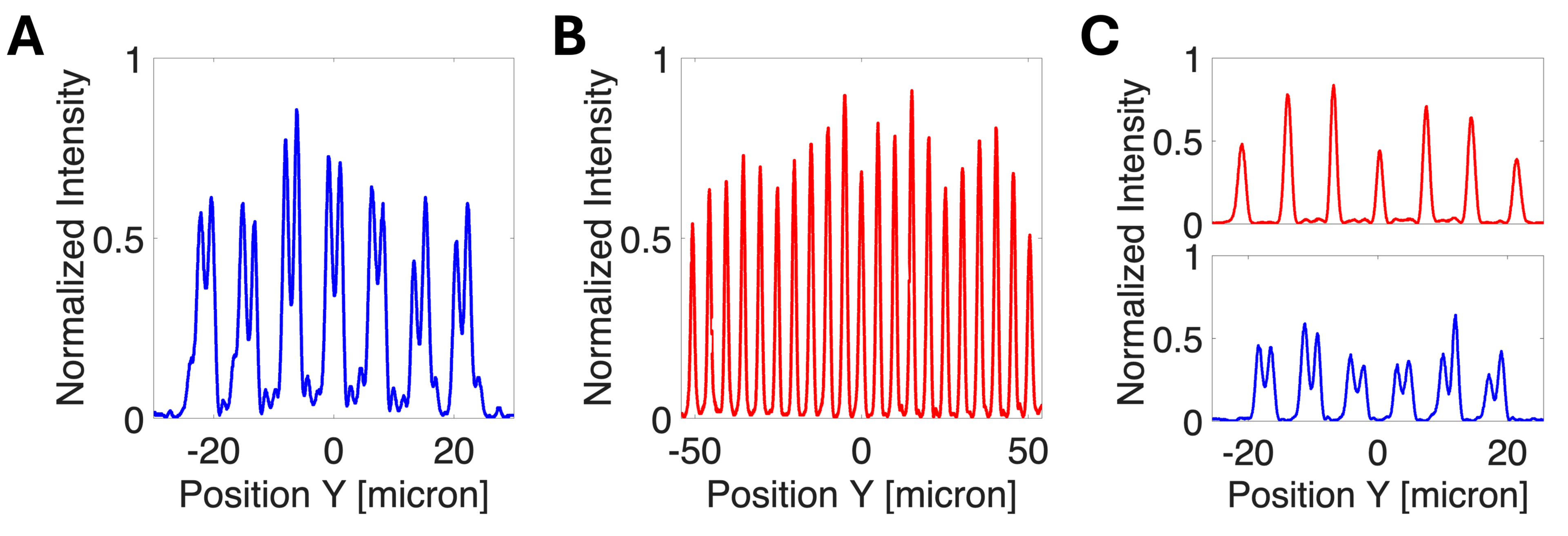} 
    \caption{ Intensity line profiles that cut through the experimental plots in Fig. 5(A-C) in the main text.  The intensity is normalized to the highest intensity across the entire focal plane. \textbf{(A)} Intensity of the dark-trap array in the focal plane at $x=0~\mu\mathrm{m}$. 
    \textbf{(B)} Intensity of the bright-trap array in the focal plane for $x=0~\mu\mathrm{m}$. 
    \textbf{(C)} Intensity of the two-species trap array at focal plane for $x=0~\mu\mathrm{m}$ (bright traps) and $x=3.5~\mu\mathrm{m}$ (dark traps).
    }
    \label{fig:si6}
\end{figure}

\newpage
\section{7. Uniformity-optimized bright-trap arrays}

\begin{figure}[H]
    \centering
    \includegraphics[width=\linewidth]{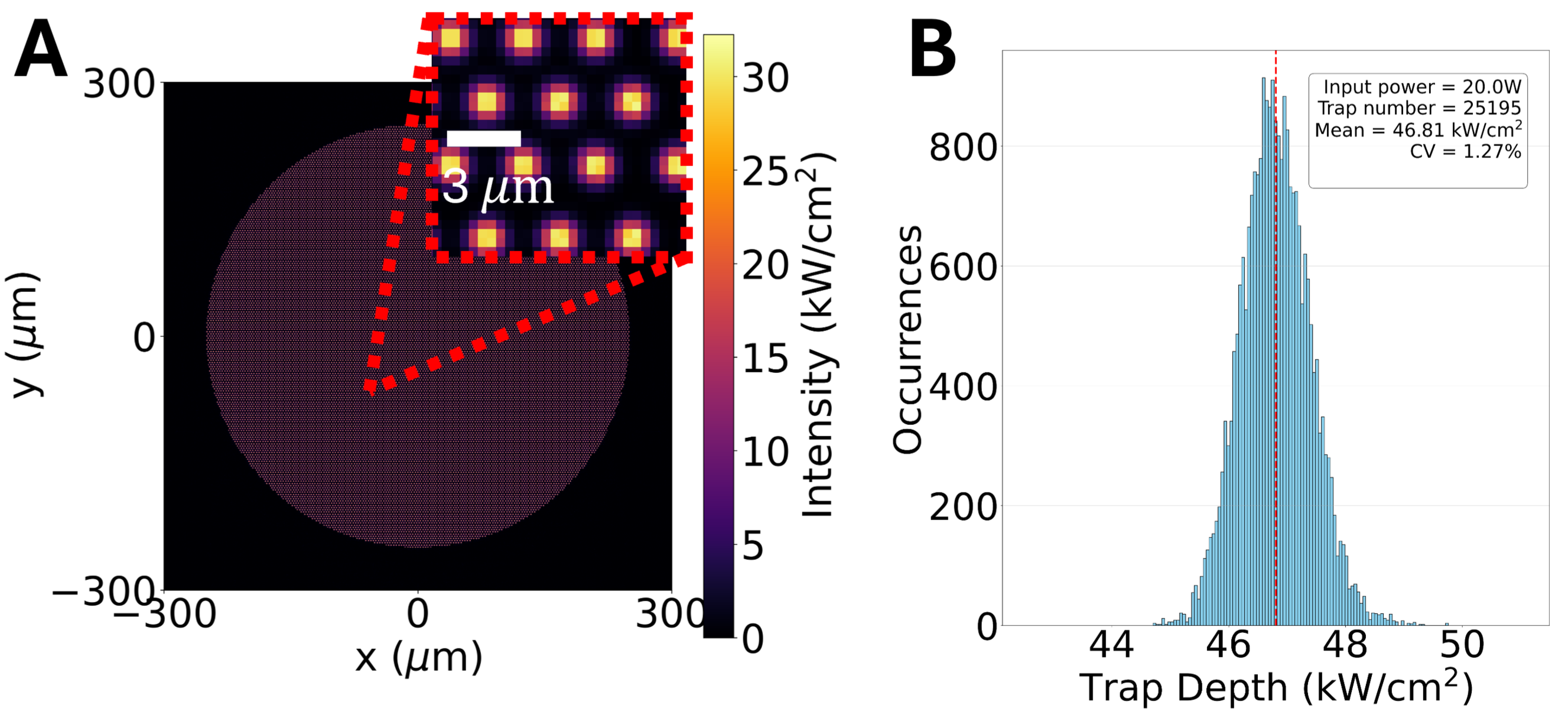} 
    \caption{
    \textbf{(A)} Simulated intensity profile at the focal plane for a hexagonally arranged circular bright-trap array with pitch of $3~\mu\mathrm{m}$ and radius of $250~\mu\mathrm{m}$. This array was optimized for trap uniformity. The inset shows a zoomed-in view of bright traps with beam waist of $0.9~\mu\mathrm{m}$. \textbf{(B)} Histogram of intensity trap depth across the entire array (25195 trap sites).
    }
    \label{fig:si7}
\end{figure}

Though this was not our singular figure of merit in Fig.~\ref{fig:figure5}(B) of the main text, metasurfaces are capable of generating large bright-trap arrays with very high uniformity. For Fig.~\ref{fig:si7}, we optimized an array with 25195 bright traps, with average trap depth of $41.81~\mathrm{kW/cm^2}$ and CV=1.27\%, assuming an input power of 20~W. In this simulation, we used the weighted G-S algorithm without any phase constraints to maximize the trap uniformity. \red{The uniformity reported here is significantly better than that of the metasurfaces designed with phase constraints in the main text. This is because, without phase constraints, the G-S algorithm has additional degrees of freedom available to optimize the trap-depth uniformity at the focal plane.}

\label{sec:si7}

\newpage
\section{8. Designing a large array of dark traps}

\begin{figure}[htbp]
    \centering
    \includegraphics[width=0.52\linewidth]{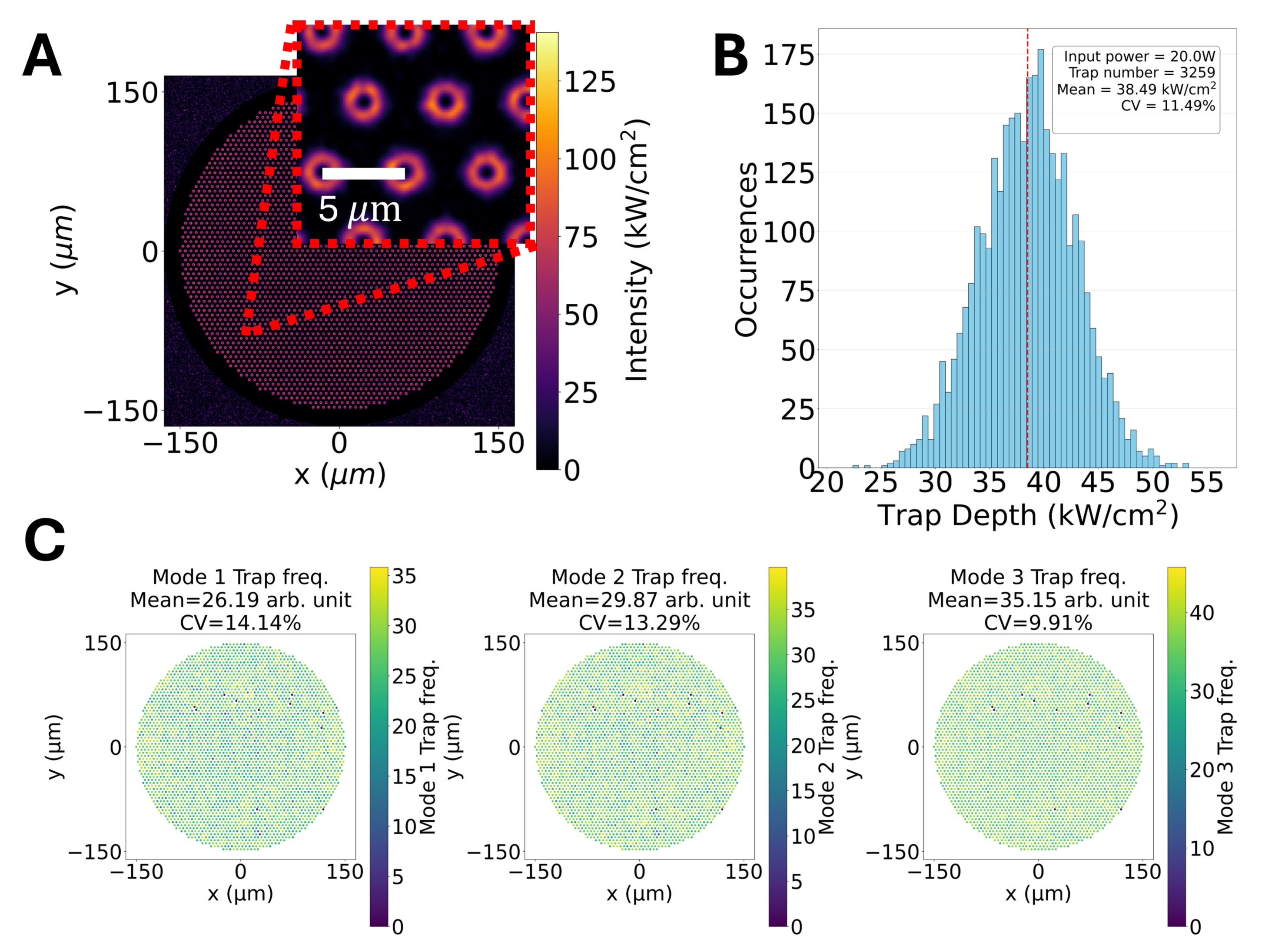} 
    \caption{
    \textbf{(A)} Simulated intensity profile at the focal plane for a hexagonally arranged circular dark-trap array with pitch of $5~\mu\mathrm{m}$ and radius of $150~\mu\mathrm{m}$. The inset shows a zoomed-in view of representative dark traps, which can be described as destructive interference between two Gaussian beams with waists $w_1=1.33~\mu\mathrm{m}$ and $w_2=0.7~\mu\mathrm{m}$.
    \textbf{(B)} Histogram of the intensity trap-depth metric across the entire array (3259 trap sites).
    \textbf{(C)} Maps of the calculated trap frequencies for the three principal motional modes at each trap site across the array.
    }
    \label{fig:si8}
\end{figure}

In this section, we demonstrate that metasurfaces are capable of generating large dark-trap arrays with moderate uniformity. While the uniformity can be further optimized, it is challenging to make dark-trap arrays with similar uniformity to bright-trap arrays (like in Fig.~\ref{fig:si7}) because dark traps are more geometrically complex than bright traps, and therefore require more degrees of freedom to form, leaving fewer degrees of freedom to optimize uniformity.  In Fig.~\ref{fig:si8}, we demonstrated 3259 dark traps with average trap depth of 38.49 kW/cm$^2$, assuming an input power of 20~W. The resulting CV = 11.49\%. \red{We further characterize the variation in trap frequency and trap-center position across the array. The three trap frequencies at each site (Fig.~\ref{fig:si8}(C)) are computed from the Hessian matrix of the local trapping potential, whose eigenvectors define the principal modes of the trap. The trap-center position, measured relative to its ideal position in the array, have standard deviations of $0.127~\mu$m, $0.14~\mu$m, and $0.069~\mu$m along $x$, $y$, and $z$, which are small compared to the array lattice constant.}

\label{sec:si8}

\newpage
\red{
\begin{quote}
      \section{9. Comparison between trapping methods}
      \label{sec:si9}

      Table~\ref{tab:si1} summarizes the optical devices that have been used to generate atom-trap tweezer arrays, including active devices (spatial light modulators, digital micromirror devices, acousto-optic deflectors) and passive devices (amplitude masks, microlens arrays, metasurfaces). For each device, we list the reframe rate, the largest experimentally demonstrated number of optical traps, the main challenges that each method is facing, and the largest number of optical tweezers that have been experimentally demonstrated. 
  \end{quote}

    \begin{table}[H]
      \ifdefined\showredtrue
        \color{customred}
        \captionsetup{font={color=customred}, labelfont={color=customred}}
      \fi
      \centering
      \caption{Comparison of optical devices used to generate atom-trap tweezer arrays.}
      \label{tab:si1}
      \footnotesize
      \renewcommand{\arraystretch}{2.0}
      \begin{tabular}{C{3cm}C{2cm}C{4cm}C{4cm}}
          \hline
          Device & Refresh rate & Challenges & Max. optical tweezers demonstrated \\
          \hline
          Spatial light modulator   & sub kHz   & Limited pixel count; diffraction efficiency        & $\sim$12,000 sites\textsuperscript{1} \\
          \hline
          Digital micromirror device & $\sim$kHz & Limited pixel count; \newline power efficiency     & 20 sites\textsuperscript{2}           \\
          \hline
          Acousto-optic deflector   & $\sim$MHz & Limited radio frequency tones          & 512 sites\textsuperscript{3}          \\
          \hline
          Amplitude mask            & static    & Power efficiency; static                          & 1,225 sites\textsuperscript{4}        \\
          \hline
          Microlens array           & static    & Filling-factor loss; static & $\sim$3,000 sites\textsuperscript{5}  \\
          \hline
          Metasurface               & static    & Static     & 360,000 sites\textsuperscript{6}      \\
          \hline
      \end{tabular}

      \vspace{1em}
      \begin{minipage}{13.8cm}
          \footnotesize
          \raggedright
          \textsuperscript{1}Manetsch et al. \textit{Nature} \textbf{2025}, \textit{647}, 60--67.\\
          \textsuperscript{2}Stuart and Kuhn. \textit{New J. Phys.} \textbf{2018}, \textit{20}, 023013.\\
          \textsuperscript{3}Singh et al. \textit{Phys. Rev. X} \textbf{2022}, \textit{12}, 011040.\\
          \textsuperscript{4}Huft et al. \textit{Phys. Rev. A} \textbf{2022}, \textit{105}, 063111.\\
          \textsuperscript{5}Pause et al. \textit{Optica} \textbf{2024}, \textit{11}, 222--226.\\
          \textsuperscript{6}Holman et al. \textit{Nature} \textbf{2026}, \textit{649}, 859--865.
      \end{minipage}
  \end{table}
  }

\end{document}